\documentclass[a4paper]{article}

\usepackage{INTERSPEECH2016}

\usepackage{graphicx}
\usepackage{amssymb,amsmath,bm}
\usepackage{textcomp} 
\usepackage{caption} 
\usepackage{subcaption} 
\usepackage{enumitem} 

\ninept

\title{Speaker-Targeted Audio-Visual Models for\\
Speech Recognition in Cocktail-Party Environments}

\makeatletter
\def\name#1{\gdef\@name{#1\\}}
\makeatother \name{{\em Guan-Lin Chao$^1$, William Chan$^1$, Ian Lane$^{1,2}$}}

\address{Carnegie Mellon University \\ $^1$Electrical and Computer Engineering, $^2$Language Technologies Institute \\
  {\small \tt \{guanlinchao,williamchan,lane\}@cmu.edu}
}

\begin{document}

  \maketitle
  \begin{abstract}
     Speech recognition in cocktail-party environments remains a significant challenge for state-of-the-art speech recognition systems, as it is extremely difficult to extract an acoustic signal of an individual speaker from a background of overlapping speech with similar frequency and temporal characteristics. We propose the use of speaker-targeted acoustic and audio-visual models for this task. We complement the acoustic features in a hybrid DNN-HMM model with information of the target speaker's identity as well as visual features from the mouth region of the target speaker. Experimentation was performed using simulated cocktail-party data generated from the GRID audio-visual corpus by overlapping two speakers's speech on a single acoustic channel. Our audio-only baseline achieved a WER of 26.3\%. The audio-visual model improved the WER to 4.4\%. Introducing speaker identity information had an even more pronounced effect, improving the WER to 3.6\%. Combining both approaches, however, did not significantly improve performance further. Our work demonstrates that speaker-targeted models can significantly improve the speech recognition in cocktail-party environments.
  \end{abstract}

  \section{Introduction}
    Automatic Speech Recognition (ASR) in cocktail-party environments aims to recognize the speech of an individual speaker from a background containing many concurrent voices, and has attracted researchers for decades \cite{bregler1994eigenlips, mcdermott2009cocktail}.
    Current ASR systems can decode clear speech well in relatively noiseless environments. However, in a cocktail-party environment, their performance is severely degraded in the presence of loud noise or interfering speech signals, especially when the acoustic signal of the speaker of interest and the background share similar frequency and temporal characteristics \cite{choi2002multichannel}. 
    Some previous approaches to this problem can be: \textit{multimodal robust features} and \textit{blind signal separation}, or a hybrid of both.
    
    In ASR systems, it is common to adapt a well-trained, general acoustic model to new users or environmental conditions.    \cite{saon2013speaker} proposed to supply speaker identity vectors, \textit{i-vectors} as input features to a deep neural network (DNN) along with acoustic features. \cite{karanasou2014adaptation} extended \cite{saon2013speaker} by factorizing i-vectors to represent speaker as well as acoustic environment. \cite{bridle1990recnorm} trained speaker-specific parameters jointly with acoustic features in an adaptive DNN-hidden Markov model (DNN-HMM) for word recognition. \cite{abdel2013fast, xue2014fast, doddipatla2014speaker} proposed training speaker-specific discriminant features (referred to as \textit{speaker codes} and \textit{bottleneck features}) for fast DNN-HMM speaker adaptation in speech recognition. \cite{abdel2013rapid} extended the speaker codes approach to convolutional neural network-HMM (CNN-HMM) systems. \cite{miao2015speaker} investigated different NN architectures of learning i-vectors for input feature mapping.
    
    Inspired by humans' ability to use other sensory information like visual cues and knowledge about the environment to recognize speech, research in audio-visual ASR has also demonstrated the advantage of using audio-visual features over audio-only features in robust speech recognition.
    The McGurk effect was introduced in \cite{mcgurk1976hearing}, which illustrates that visual information can affect human's interpretation of audio signals. In \cite{bregler1994eigenlips}, low dimensional lip movement vectors, \textit{eigenlips}, were used to complement acoustic features for ASR.
    In \cite{nefian2002dynamic}, generalized versions of HMMs, factorial HMM and the coupled HMM, were used to fuse auditory and visual information, in which the HMM parameters were able to be trained with dynamic Bayesian networks. In \cite{ngiam2011multimodal}, the authors proposed a DNN-based approach to learning multimodal features and a shared representation between modalities. In \cite{mroueh2015deep}, the authors presented a deep neural network that used a bilinear softmax layer to account for class specific correlations between modalities.
    In \cite{noda2015audio}, a deep learning architecture with multi-stream HMM model was proposed. Using noise-robust acoustic features extracted by autoencoders and mouth region of interest (ROI) image features extracted by CNNs, this approach achieved higher word recognition rate than the use of non-denoised features or normal HMMs. \cite{biswas2016multiple} proposed an active appearance model-based approach to extracting visual features of jaw and lip ROI on four image streams which were then combined with acoustic features for in-car audio-visual ASR.
    Traditional cocktail-party ASR methods suggest performing blind signal separation prior to auditory speech recognition of individual signals. Blind signal separation aims at estimating multiple unknown sources from the sensor signals. When there is only a single-channel signal available, source separation on the cocktail-party problem becomes even more difficult \cite{schmidt2006single}. A main assumption in the signal separation is that speech signals from different sources are statistically independent \cite{choi2002multichannel}. Another common assumption of signal separation is that all the sources have zero-mean and unit variance for the convenience of performing Independent Component Analysis \cite{comon1994independent, jang2002probabilistic}.
    However, these two assumptions are not always correct in practice. Therefore, we try to lift these assumptions by directly recognizing single-channel signals of overlapping speech in this work.
    
    In this paper, we propose a speaker-targeted audio-visual ASR model of multi-speaker acoustic input signals in cocktail-party environments without the use of blind signal separation. With the term \textit{speaker-targeted} model, we refer to a speaker-independent model with speaker identity information input. We complement the acoustic features with information of the target speaker's identity in embeddings similar to i-vectors in \cite{saon2013speaker}, along with raw pixels of the target speaker's mouth ROI images, to supply multimodal input features to a hybrid DNN-HMM for speech recognition in cocktail-party environments.

  \section{Model}
    In this work, we focus on the cocktail-party problem with overlapping speech from two speakers. We approach this problem using DNN acoustic models with different combinations of additional modalities: visual features and speaker identity information. The acoustic features are filterbank features extracted from the audio signals where two speakers' speech is mixed on a single acoustic channel. The visual features are raw pixel values of the mouth ROI images of the target speaker whose speech the system is expected to recognize. The speaker identity information is represented by the target speaker's ID-embedding.
    Details about feature extraction are described in the Experiments section.
  
    DNN acoustic models have been widely and successfully used in ASR \cite{hinton-ieeespm-2012}. Let $\mathbf x$ be a window of acoustic frames (i.e., context of filterbanks), the standard DNN acoustic models model the posterior probability:
    \begin{align}
        p(y | \mathbf x) = \mathrm{DNN}(\mathbf x) 
    \end{align}
    where $y$ is a phoneme label or alignment (i.e., from GMM-HMM) and $\mathrm{DNN}$ is a deep neural network with softmax outputs. The $\mathrm{DNN}$ is typically trained to maximize the log probability of the phoneme alignment or minimize the cross-entropy error. However, this optimization problem is difficult when $\mathbf x = \mathbf x_1 + \mathbf x_2$ is a superposition of two signals $\mathbf x_1$ and $\mathbf x_2$ (i.e., cocktail party).
    
    In this work, we extend the previous traditional DNN acoustic models to leverage additional information in order to model our phonemes.
    By leveraging combinations of the visual features and speaker identity information, the standard DNN acoustic model is extended to have multimodal inputs.
    
    We train the DNN acoustic models with four possible combinations of input features, audio-only, audio-visual, speaker-targeted audio-only, and speaker-targeted audio-visual,
    in two steps: \textit{speaker-independent models training} followed by \textit{speaker-targeted models training}. The details of the two steps are described in the following sub-sections.
    
    \subsection{Two-Speaker Speaker-Independent Models}
\begin{figure}[t]
        \centering
        \includegraphics[width=0.5\linewidth]{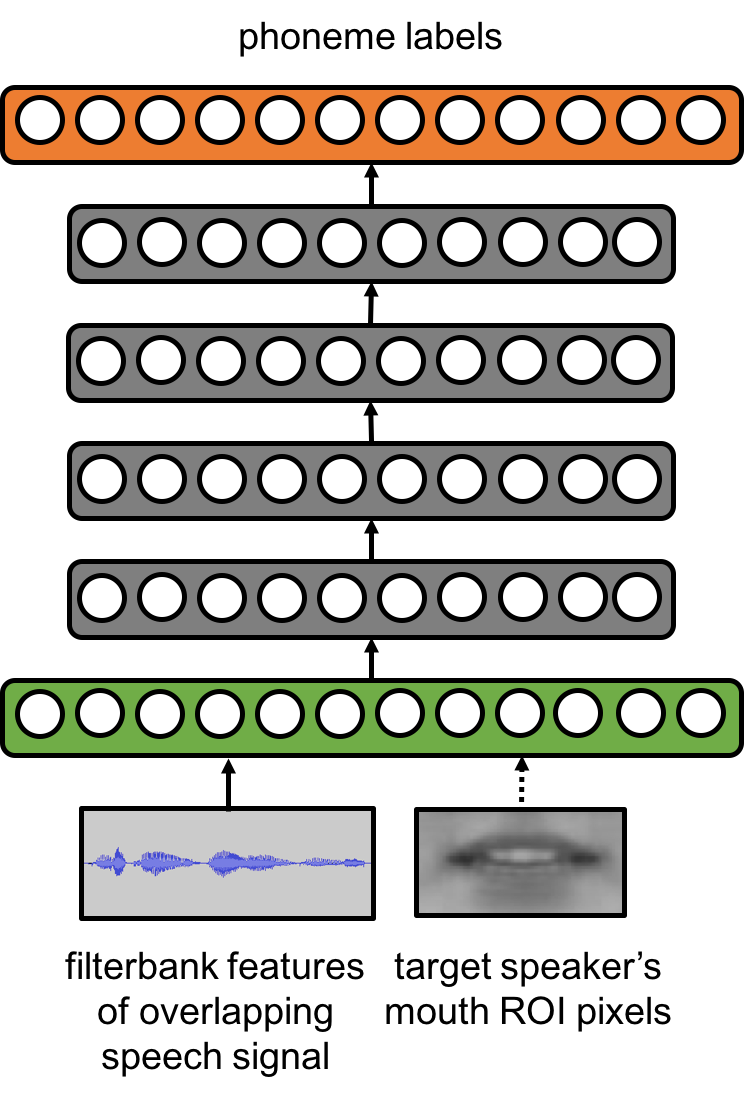}
        \caption{{\it Speaker-Independent Models. This figure illustrates a DNN architecture, where the phoneme labels are modeled in the output layer. The arrow connecting the visual features (mouth ROI pixels) and the input layer is a dashed arrow. The speaker-independent audio-only model is illustrated without the dashed arrow. The speaker-independent audio-visual model is illustrated with the dashed arrow, where the acoustic features (filterbank features) and video features are concatenated as DNN inputs.}}
        \label{fig:avmodel}
      \end{figure}

      First, we leverage the visual information in conjunction with the acoustic features.
      The standard DNN acoustic model:
      \begin{align}
        p(y | \mathbf x) &= \mathrm{DNN_A}(\mathbf x) \label{eq:model_A}
      \end{align}
      with additional input of visual features becomes:
      \begin{align}
        p(y | \mathbf x, \mathbf w) &= \mathrm{DNN_{AV}}(\mathbf x, \mathbf w) \label{eq:model_AV}
      \end{align}
      where $\mathbf w$ are the visual features.
      In this step, speaker-independent audio-only model $\mathrm{DNN_A}$ and audio-visual model $\mathrm{DNN_{AV}}$ are trained for the two-speaker cocktail-party problems. The acoustic and visual features are concatenated directly as DNN inputs for the audio-visual model. The speaker-independent models are illustrated in Figure \ref{fig:avmodel}, where the figure without the dashed arrow represents the audio-only model, and the figure with the dashed arrow represents the audio-visual model. 
    
    \subsection{Two-Speaker Speaker-Targeted Models}
\begin{figure*}[h!]
        \centering
        \begin{subfigure}[t]{0.3\textwidth}
        \centering\captionsetup{width=.9\linewidth}%
          \includegraphics[width=\linewidth]{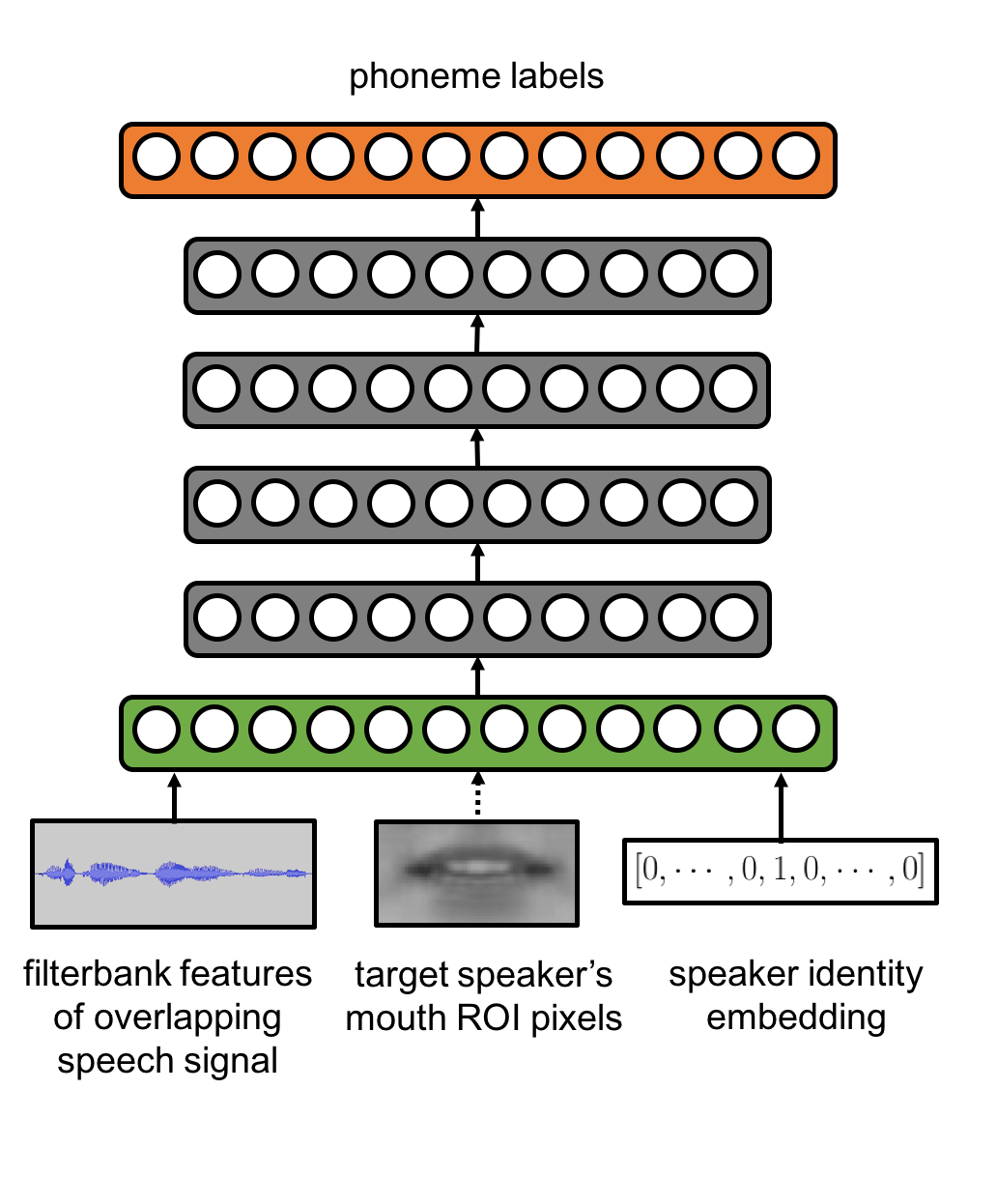}
          \caption{{\it concatenating the speaker identity directly with audio-only and audio-visual features}}
          \label{fig:fusionA}
        \end{subfigure}
        \begin{subfigure}[t]{0.3\textwidth}
        \centering\captionsetup{width=.9\linewidth}%
          \includegraphics[width=\linewidth]{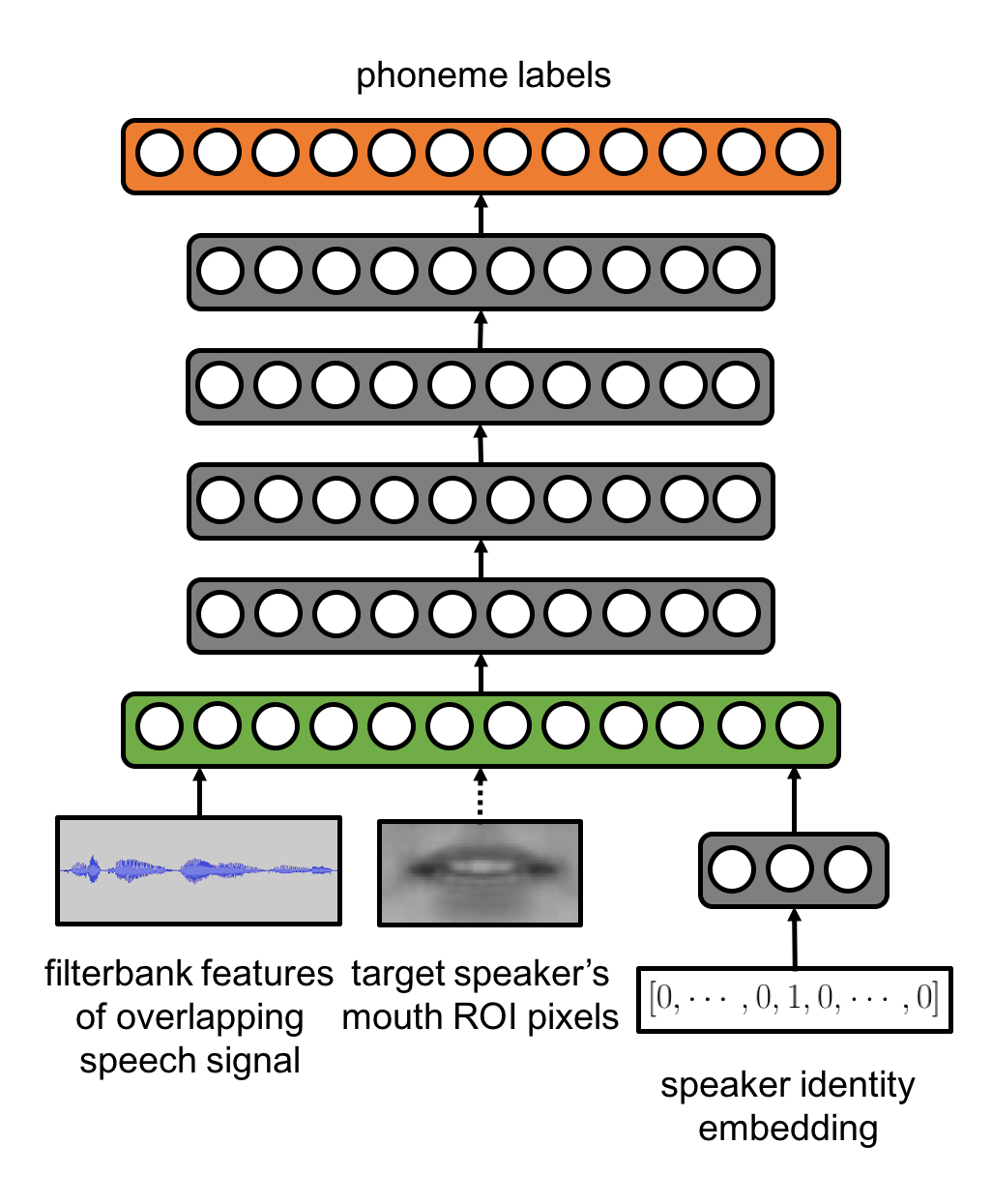}
          \caption{{\it mapping speaker identity into a compact but presumably more discriminative embedding and then concatenating the compact embedding with audio-only and audio-visual features}}
          \label{fig:fusionB}
        \end{subfigure}
        \begin{subfigure}[t]{0.3\textwidth}
        \centering\captionsetup{width=.9\linewidth}%
          \includegraphics[width=\linewidth]{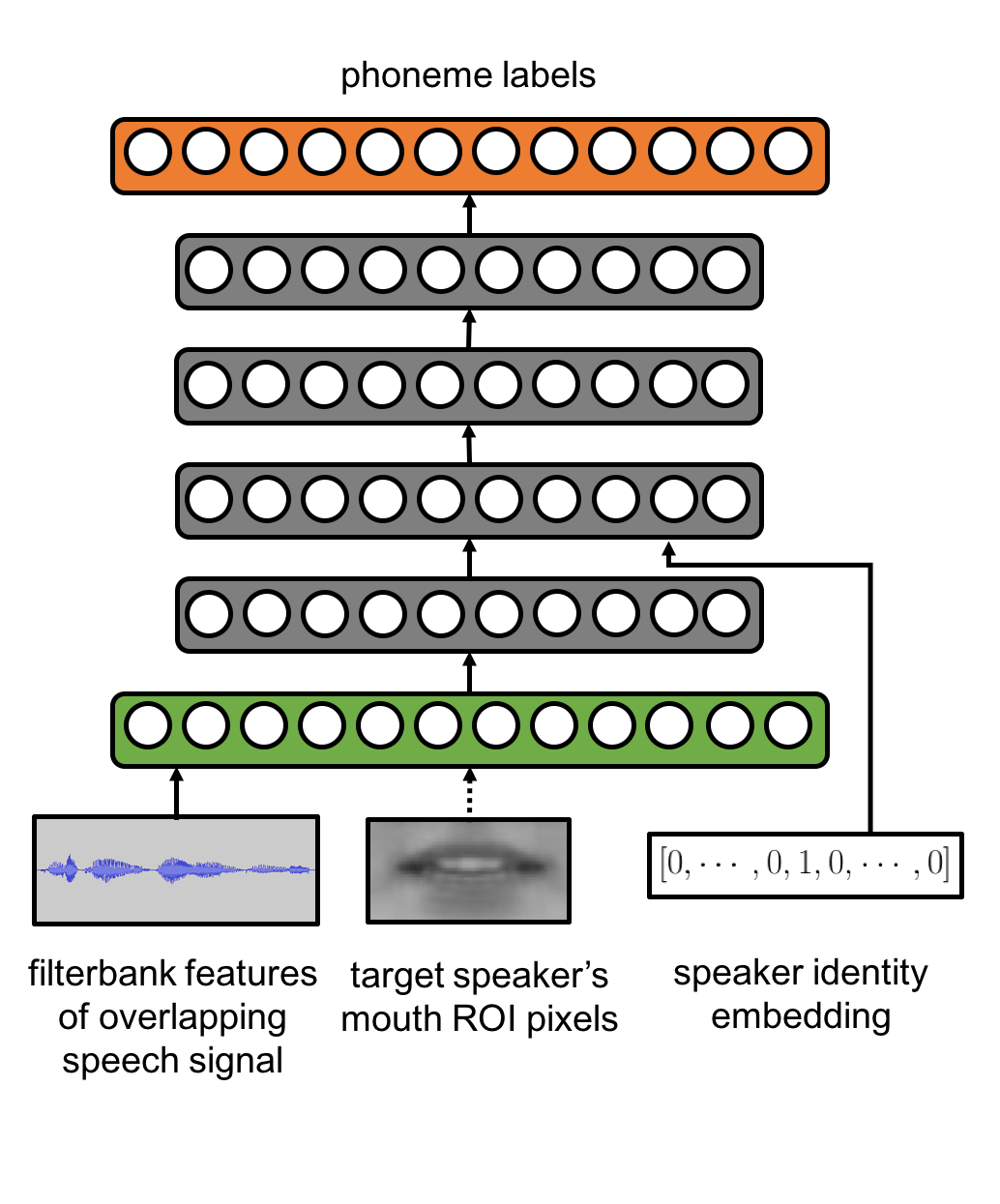}
          \caption{{\it connecting the speaker identity to a later layer than audio-only and audio-visual features}}
          \label{fig:fusionC}
        \end{subfigure}
        \caption{{\it Three Variants of Speaker-Targeted Models. These figures illustrate three fusion techniques of audio-visual features with speaker identity information in a DNN architecture, where the phoneme labels are modeled in the output layer. The arrows connecting the visual features
        and the input layers are dashed arrows. The speaker-targeted audio-only models are illustrated without the dashed arrows. The speaker-targeted audio-visual models are illustrated with the dashed arrows, where the acoustic 
        and video features are concatenated as DNN inputs.}}
        \label{fig:fusions}
      \end{figure*}

      Secondly, we try to leverage the speaker identity information to extend the previous models, $\mathrm{DNN_{A}}$ and $\mathrm{DNN_{AV}}$. 
      $\mathrm{DNN_{A}}$ is extended to:
      \begin{align}
        p(y | \mathbf x, \mathbf z) &= \mathrm{DNN_{AI}}(\mathbf x, \mathbf z) \label{eq:model_AI}
      \end{align}
      and $\mathrm{DNN_{AV}}$ is extended to:
      \begin{align}
        p(y | \mathbf x, \mathbf w, \mathbf z) &= \mathrm{DNN_{AVI}}(\mathbf x, \mathbf w, \mathbf z) \label{eq:model_AVI}
      \end{align}
      where $\mathbf z$ are the speaker identity information.
      In this step, we adapt the audio-only and audio-visual speaker-independent models to speaker-targeted models respectively (i.e., from $\mathrm{DNN_{A}}$ to $\mathrm{DNN_{AI}}$ and from $\mathrm{DNN_{AV}}$ to $\mathrm{DNN_{AVI}}$) by hinting the network which target speaker to attend to by supplying speaker identity information as input.
      The speaker identity information is represented by an embedding that corresponds to the target speaker's ID. 
      We investigate three ways to fuse the audio-visual features with the speaker identity information:
      \begin{enumerate}[label=(\Alph*)]
          \item Concatenating the speaker identity directly with audio-only and audio-visual features.
          \item Mapping speaker identity into a compact but presumably more discriminative embedding and then concatenating the compact embedding with audio-only and audio-visual features.
          \item Connecting the speaker identity to a later layer than audio-only and audio-visual features.
      \end{enumerate}
      The three fusion techniques introduce the three variants (A), (B) and (C) of both the speaker-targeted models $\mathrm{DNN_{AI}}$ and $\mathrm{DNN_{AVI}}$. The speaker-targeted models of the three invariants are shown in Figure \ref{fig:fusions}, where the figures without the dashed arrow represent the audio-only models, and the figures with the dashed arrow represent the audio-visual models.
      
      \begin{figure*}[t]
          \centering
          \includegraphics[width=\linewidth]{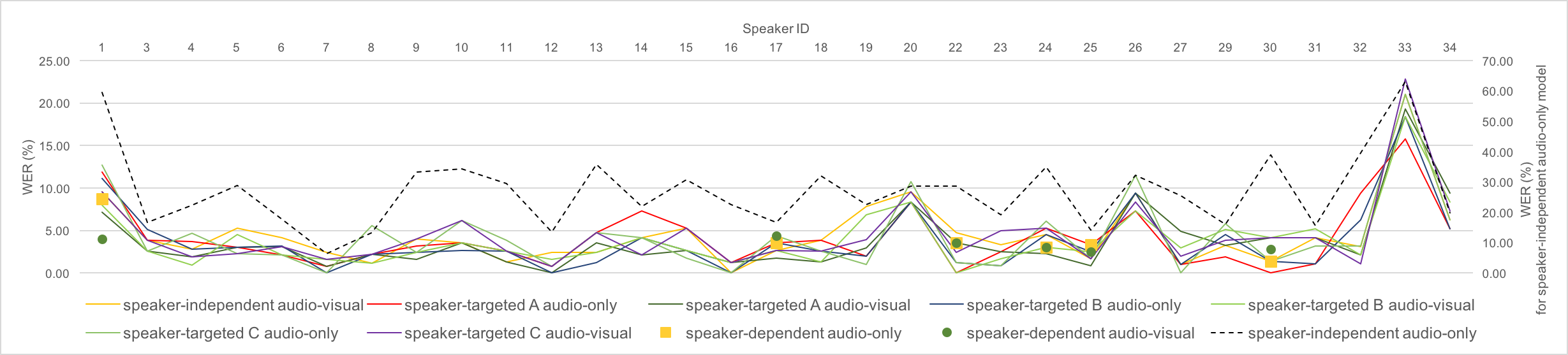}
          \caption{{\it WER Comparisons of Two-Speaker models for Individual Speakers. WER of two-speaker models for individual speakers are illustrated. The dashed line is plotted on the right vertical axis which represents the speaker-independent audio-only model. The solid lines and markers are plotted on the left vertical axis. Speaker-dependent models for speaker 1, 17, 22, 24, 25 and 30 are plotted in markers. The chart demonstrates a similar trend between different models' performance on individual speakers.}}
          \label{fig:wer_breakdown}
      \end{figure*}
      
      Moreover, we train single-speaker speaker-independent models in comparison with the two-speaker speaker-independent models.
      We also train 6 randomly-selected speaker's speaker-dependent models (adapted from speaker-independent models as well) to compare with the speaker-targeted models.

  \section{Experiments}
    \subsection{Dataset}
      The GRID corpus \cite{cooke2006audio} is a multi-speaker audio-visual corpus. This corpus consists of high-quality audio and video recordings of 34 speakers in quiet and low-noise conditions. Each of the speakers read 1000 sentences which are simple six-word commands obeying the following syntax: 
      
      \textit{\$command \$color \$preposition \$letter \$digit \$adverb}
      
      We use the utterances of 31 speakers (16 males and 15 females) from the GRID corpus, excluding speaker 2, 21 and 28 and part of the utterances of the remaining 31 speakers due to the availability of mouth ROI image data.
      In the one-speaker datasets, there are 15395 utterances in the training set, 548 in the validation set, and 540 in the testing set, following the convention of CHiME Challenge \cite{barker2013pascal}. The GRID corpus utterances that don't belong to the one-speaker datasets are termed background utterance set. To simulate the overlapping speech audios for the two-speaker datasets, we mix the target speaker and a background speaker's utterances with equal weights on a single acoustic channel using SoX software \cite{sox}. The background speaker's utterances are randomly selected from the background utterance set excluding the utterances of the target speaker. The resulting mixed audio's length is as long as the length of the target speaker's utterance.
    \subsection{Feature Extraction}
    
      \subsubsection{Audio Features}
        Log-mel filterbank features with 40 bins are extracted, and a context of $\pm 5$ frames was used for audio input features (i.e., $440 = 40 * (5 + 1 + 5)$ dimensions per acoustic feature $\mathbf x$).
        
      \subsubsection{Visual Features}
        We use target speaker's mouth ROI images' pixel values as visual features. The facial landmarks are first extracted by IntraFace software \cite{de2015intraface}, 
        and each video frame is cropped into a 60 pixel * 30 pixel mouth ROI image \cite{zehngut2015audio} according to the mouth region landmarks (i.e., $1800 = 60 * 30$ dimensions per visual feature $\mathbf w$). The gray-scale pixel values are then concatenated with audio features to form audio-visual features.
        
      \subsubsection{Speaker Identity Information}
        Speaker identity information is represented by the target speaker's ID-embedding, which is simply a one-hot vector of thirty-three 0s and a single 1, $[0, \cdots, 0, 1, 0, \cdots, 0]$, in which the entry of 1 corresponds to the target speaker's ID (i.e., $34$ dimensions per speaker identity embedding $\mathbf z$).
        
    \subsection{Acoustic Model}
      Here we describe the architecture of our DNNs.
      The number of hidden layers for audio-only models and speaker-independent audio-visual models is 4, while it is 5 for speaker-targeted and speaker-dependent audio-visual models. Each hidden layer contains 2048 nodes. Rectified linear function (ReLU) is used for activation in each hidden layer. The output layer has a softmax of 2371 phoneme labels. We use stochastic gradient descent with a batch size of 128 frames and a learning rate of 0.01.

    \subsection{Results}
      \begin{table}[th]
        \caption{\label{tab:wer_singlespeaker} {\it WER Comparisons of Single-Speaker Models}}
        \centerline{
          \begin{tabular}{| c | c | c |}
            \hline
            \multicolumn{1}{|c|}{} &
            \multicolumn{1}{|c|}{audio-only} & 
            \multicolumn{1}{|c|}{audio-visual} \\
            \hline \hline
            speaker-independent & $0.3\%$ & $0.4\%$ \\
            \hline
          \end{tabular}
        }
      \end{table}
      \begin{table}[th]
        \caption{\label{tab:wer_twospeaker} {\it WER Comparisons of Two-Speaker Models}}
        \centerline{
          \begin{tabular}{| c | c | c |}
            \hline
            \multicolumn{1}{|c|}{} &
            \multicolumn{1}{|c|}{audio-only} & 
            \multicolumn{1}{|c|}{audio-visual} \\
            \hline \hline
            speaker-independent & $26.3\%$ & $4.4\%$ \\
            \hline
            speaker-targeted A & $4.0\%$ & $3.6\%$ \\ 
            \hline
            speaker-targeted B & $3.6\%$ & $3.9\%$ \\ 
            \hline
            speaker-targeted C & $4.4\%$ & $4.4\%$ \\ 
            \hline
            speaker-dependent & $3.9\%$ & $3.4\%$ \\ 
            \hline
          \end{tabular}
        }
      \end{table}
      
      The aforementioned single-speaker models are used to decode the single-speaker testing dataset, while the two-speaker models are used to decode the two-speaker testing dataset. Table \ref{tab:wer_singlespeaker} shows the WER of single-speaker models. Table \ref{tab:wer_twospeaker} shows the WER of two-speaker models.
      The audio-only baseline for two-speaker cocktail-party problem is 26.3\%. The results of speaker-independent models for single-speaker and two-speaker suggest that automatic speech recognizers' performance degrades severely in cocktail-party environments compared to low-noise conditions. It is also demonstrated that the introduction of visual information to acoustic features can reduce WER significantly in cocktail-party environments, improving the WER to 4.4\%, although it may not help when the environmental noise is low. WER comparisons between two-speaker's audio-only speaker-independent and speaker-targeted models suggest that using speaker identity information in conjunction with acoustic features achieves a better improvement on WER, reducing WER up to 3.6\%. 
      
      The results of two-speaker's speaker-targeted models A, B, and C suggest a weak tendency that providing speaker information in earlier layers of the network seems to have advantage. WER comparisons between two-speaker speaker-dependent and speaker-targeted models 
      suggest an intuitive result that a speaker-dependent ASR system which is optimized for one specific speaker performs better than a speaker-targeted ASR system which is optimized for multiple speakers simultaneously.
      We also find the introduction of visual information improves the WER of speaker-dependent acoustic models while it doesn't improve the speaker-targeted acoustic models. We subscribe this finding to the limitation of the capacity of the neural network architecture that we use for both models, that it is able to optimize for one specific speaker's visual information in a speaker-dependent model, but not powerful enough to learn a unified optimization for all 31 speakers' visual information in a single speaker-targeted model. Figure \ref{fig:wer_breakdown} illustrates the WER of the individual speakers. A similar trend between different models' performance on individual speakers is demonstrated.

  \section{Conclusions}
    A speaker-targeted audio-visual DNN-HMM model for speech recognition in cocktail-party environments is proposed in this work. Different combinations of acoustic and visual features and speaker identity information as DNN inputs are presented. Experimental results suggest that the audio-visual model achieves significant improvement over the audio-only model. Introducing speaker identity information introduces an even more pronounced improvement. Combining both approaches, however, does not significantly improve performance further.
    
    Future work will aim to investigate better representations in multimodal data space to incorporate audio, visual and speaker identity information with the objective to improve the speech recognition performance in cocktail-party environments. More complex architectures can be explored such as CNNs for modeling image structures and recurrent neural networks (RNNs) and Long-Short Term Memory (LSTM) models for modeling variable time-sequence inputs in order to achieve a better unified optimization for the speaker-targeted audio-visual models.

  \section{Acknowledgements}
    We would like to thank Niv Zehngut for preparing the image dataset, Srikanth Kallakuri for technical support, Benjamin Elizalde and Akshay Chandrashekaran for proofreading and suggestions.
    
  \newpage
  \eightpt
  \bibliographystyle{IEEEtran}

  \bibliography{guanlinc}

\end{document}